\documentclass[aip,reprint]{revtex4-2}
\usepackage{graphicx,amssymb,amsmath}
\usepackage{color}
\usepackage[dvipsnames]{xcolor}
\begin{document}
\title{Ostwald ripening of aqueous microbubble solutions} 
\author{Sota Inoue}
\author{Yasuyuki Kimura}
\affiliation{Department of Physics, Kyushu University, Fukuoka 819-0395, Japan}
\author{Yuki Uematsu}
\email{uematsu@phys.kyutech.ac.jp}
\affiliation{Department of Physics and Information Technology, Kyushu Institute of Technology, Iizuka 820-8502, Japan}
\affiliation{PRESTO, Japan Science and Technology Agency, 4-1-8 Honcho, Kawaguchi, Saitama 332-0012, Japan}
\date{\today}


\begin{abstract}
Bubble solutions are of growing interest because of various technological applications in surface cleaning, water treatment, and agriculture. 
However, their physicochemical properties such as the stability and interfacial charge of bubbles are not fully understood yet. 
In this study, the kinetics of radii in aqueous microbubble solutions are experimentally investigated, and the results are discussed in the context of Ostwald ripening. The obtained distributions of bubble radii scaled by mean radius and total number were found to be time-independent during the observation period. 
Image analysis of radii kinetics revealed that the average growth and shrinkage speed of each bubble is governed by diffusion-limited Ostwald ripening, and the kinetic coefficient calculated using the available physicochemical constants in literature quantitatively agrees with the experimental data.
Furthermore, the cube of mean radius and mean volume exhibit a linear time evolution in agreement with the Lifshitz-Slezov-Wagner (LSW) theory.
The coefficients are slightly larger than those predicted using the LSW theory, which can be qualitatively explained by the effect of finite volume fraction.
Finally, the slow down and pinning of radius in the shrinkage dynamics of small microbubbles are discussed in detail.
\end{abstract}

\maketitle

\section{Introduction}

Aqueous bubble solutions are of growing interest \cite{Craig2016,Swenson2018,Hu2021} because they have potential technological applications in surface cleaning, water treatment, oyster cultures, and cultivation of vegetables despite their physicochemical and hydrodynamic simplicity. 
The stability of a single isolated bubble in water has been studied theoretically \cite{Rayleigh1917,Plesset1949,Epstein1950,Plesset1982}.
For example, the lifetime of a 100nm-radius bubble is calculated to be less than $1\,\mu$s due to the dissolution of gas molecules to the surrounding water driven by a strong Laplace pressure of approximately $1\,$MPa \cite{Rayleigh1917,Plesset1949,Epstein1950,Plesset1982,Duncan_2004,Kentish_2006}.
In contrast, a stable bulk nanobubble solution without any surfactants has been repeatedly reported using ultrasonic irradiation \cite{Kim2000}, decompression of pressurized gas/water mixture \cite{Ushikubo2010}, alcohol/water exchange \cite{Qiu2017}, and shrinkage of microbubbles \cite{Jin2019}. 
These nanobubbles have radii of less than $1\,\mu$m and negative zeta potential \cite{Kim2000,Ushikubo2010}, and are sufficiently stable through various experimental characterizations.
Most theoretical studies have explained this anomalous stability by introducing the impurity film at the interface \cite{Ducker2009} or considering the electrostatic stress effect \cite{Tan2020,Satpute_2021} focused on a single isolated bubble.
However, in practice, micro- and nanobubbles are usually used as aqueous solutions, not as isolated bubbles, and the stability of dispersions may differ from that of a single isolated bubble. 

In this context, Ostwald ripening of surface bubbles was studied by an analytic theory \cite{Dollet2016} and numerical simulations \cite{Michelin2018,Zhu2018}.   
Ostwald ripening is a characteristic coarsening process of multiple particles, in which small particles dissolve rapidly, whereas large particles grow slowly because of the Gibbs-Thomson effect \cite{Ostwald1897}.
This process is a kind of phase transition dynamics in quenched systems \cite{OnukiBook,Tanaka2015}, and it is ubiquitous in nature, for example, in supersaturated solutions of crystals \cite{Ostwald1897}, metal particles in an alloy \cite{Rastogi1971,Jack1972,Hirata1977,Seno1983}, mineral particles in its vapor \cite{Lautze2011}, and oil-in-water emulsions \cite{Kabalnov1987, McClements1999, Abe2002,Taylor2003,McClements2006, Dungan2010}.
Half a century after Ostwald's work \cite{Ostwald1897}, the quantitative theory on the asymptotic behavior of coarsening was established using the mean-field kinetic theory \cite{Lifshitz1959, Lifshitz1961, Wagner1961} called Lifshitz-Slezov-Wagner (LSW) theory.
This theory demonstrated that the scaled distribution of particle radii has a universal shape in the limit of infinite time. Moreover,  the cube of the mean radius and the inverse of number density obeys a power law as $t^\alpha$ in this limit \cite{Lifshitz1959,Wagner1961}. 
The exponent is $\alpha=1$ for diffusion-limited Ostwald ripening, and another exponent is predicted for different mass-transfer mechanisms of particle coarsening \cite{Alexandrov_2017}.
It is known that experimental scaled distributions are broader than the universal scaled distribution derived by the LSW theory \cite{Voorhees1985, Marder1985, Marder1987, Baldan_2002}.  Several theories have been proposed to explain this discrepancy between the LSW theory and experiments, including the finite-volume fraction effect of coarsening phase \cite{Ardell_1972, Brailsford_1979, Tsumuraya_1983, Marqusee1984, Voorhees1985, Marder1985, Enomoto_1986, Marder1987, Yao_1993, Yao1994, Baldan_2002}. A major explanation for this discrepancy is the omission of long-range interaction of diffusion fields in the LSW theory. However, a quantitative description is still under debate.

Although the Ostwald ripening of gas bubbles have been well studied by theory \cite{Schmeizer1987,Slezov2005,Dollet2016} and simulations \cite{Watanabe2014,Watanabe2016,Michelin2018,Zhu2018} so far, few studies experimentally investigated the Ostwald ripening of bubbles \cite{Rocha2018arXiv,Rocha2018PhD}.
In Refs.~\citenum{Rocha2018arXiv} and \citenum{Rocha2018PhD}, they studied the Ostwald ripening of bubbles in glycerin/water mixture and compared the results with the LSW theory. 
However, they could not discuss quantitative aspects of coarsening dynamics because the physicochemical constants of the glycerin/water mixture were unavailable.
Moreover, they did not deduce the shrinkage and growth rate of individual bubbles from image analysis.
Thus, an experimental study on quantitative examination of the theories of Ostwald ripening in aqueous microbubble solutions is still lacking. 
In this paper, we performed experiments on the kinetics of microbubble radii and examined the theory of Ostwald ripening in aqueous microbubble solutions.
Using microscope observations and image analysis, we experimentally deduced the kinetic equations and obtained the time evolution of distribution function and mean radius.
Finally, we present a quantitative analysis by comparing our experimental data with the reported theories of Ostwald ripening using literature values for physicochemical constants.
Our results clearly demonstrate that aqueous microbubble solutions confined in a narrow rectangular glass capillary exhibit diffusion-limited Ostwald ripening.
Furthermore, the average growth rate of each bubble radius is equal to the LSW theory without any fitting parameters. 
We believe that these findings reveal new insights into the stability of bubbles and the general theory of Ostwald ripening. 
   
\section{Experimental}

\begin{figure}
	\includegraphics[width=8.4cm]{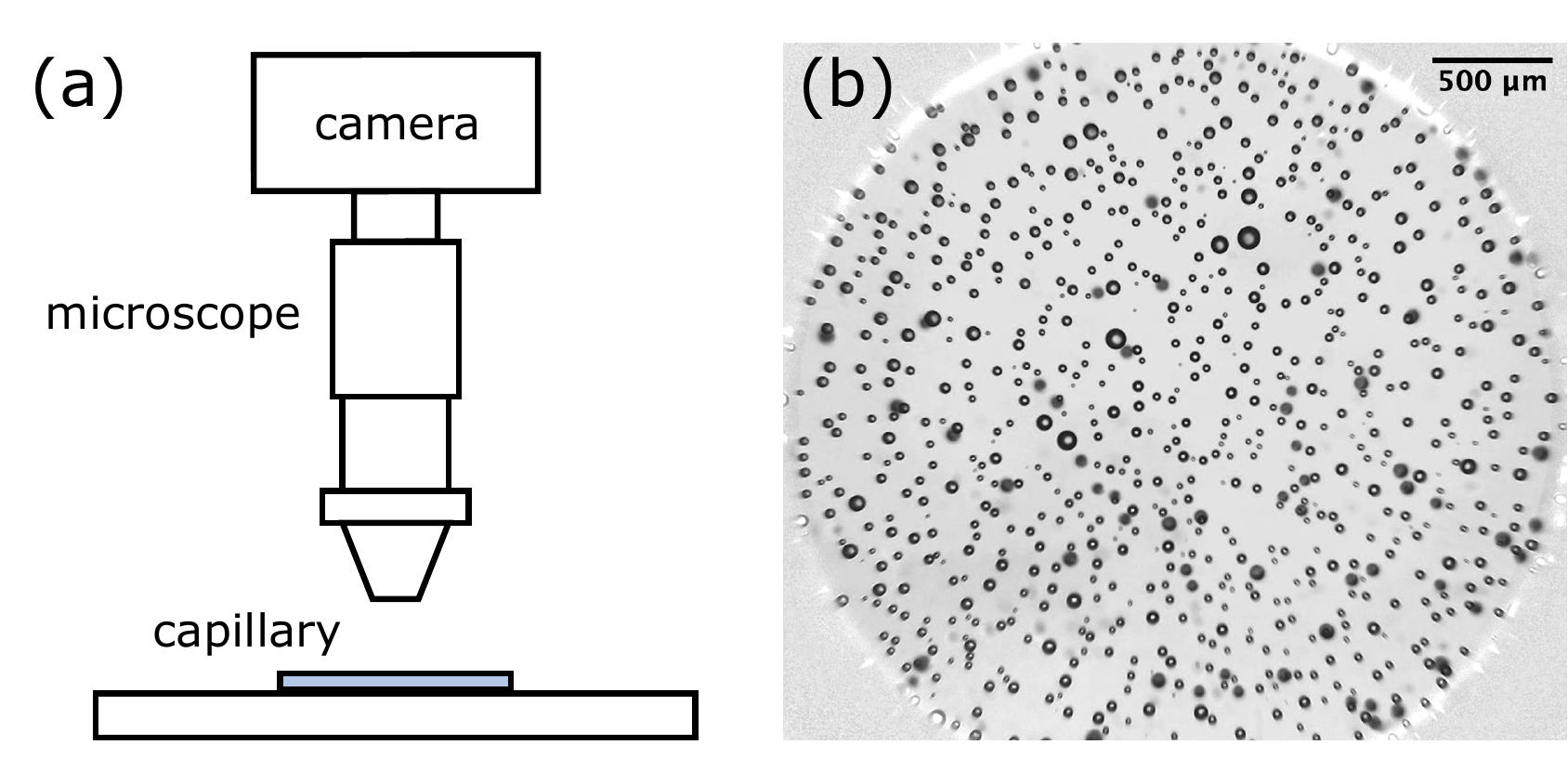}
\caption{
	(a) Schematic of the experimental setup.
	(b) Typical snapshot of microscope observation at $t=1\,$min.
}
\label{fig:0}
\end{figure}

A commercial pressurized-dissolution microbubble generator (Aura Tec, OM4-MDG-045) was used to produce microbubble solutions. 
This generator comprised inlet filters for water and gas, a pump, a pressurized tank, and an outlet nozzle.
Water was pumped into the tank, and air dissolved into the water under a gauge pressure of $0.25\,$MPa.
Next, the air-saturated water was sent to the outlet nozzle and released under atmospheric pressure to generate microbubbles.  
Approximately $3\,$L of an aqueous $10\,$mM potassium chloride solution was used to produce microbubble solutions.
Deionized water and potassium chloride (FUJIFILM Wako, $\ge99.9\%$) were used without further purification.
Few minutes after circulating the microbubble solution, the solution was injected into a rectangular glass capillary (VitroCom, 3536-050 and 4806-050, length: $50\,$mm, width: $6\,$mm, and inner thickness: $d=0.3\,$mm and $0.6\,$mm).
The reference time was set to $t=0$ at the injection of the solution.
Then, the capillary was quickly set on the microscope (Olympus, BX50), and observed by a bright-field microscope with an objective lens of $\times 4$ and a relay lens of $\times 0.5$.
The images were captured by a camera (Motic, Moticam 2000) for a total duration of $90\,$min at $1\,$min intervals until ($t=1$, $2,\cdots 90\,$min) and after that $30\,$min intervals from $t=120\,$min until $600\,$min.
The $10\,$fps images were captured by another camera (The Imaging Source, DMK33UX174) with an objective lens of $\times 20$.
During the experiments, the capillary was open to the external atmosphere.
Fig.~\ref{fig:0}a shows a schematic of the experimental setup.

Fig.~\ref{fig:0}b is a typical snapshot of observed bubbles at $t=1\,$min.
The optical imaging revealed that the bubbles were on both the top and bottom capillary surface.
However, the number of bubbles is typically much less on the bottom compared to the top surface. 
We focused on the bubbles attached to the top surface. 

The captured images were first processed by an ImageJ macro to obtain the bubble lateral positions and their radius $R_i$.
The macro is composed of the following functions: subtract background, threshold, convert to mask, fill holes, watershed, and analyze particles.
Then, we used a homemade python script to track the bubbles using maximum speed of the center of bubbles $30\,\mu$m/frame.
When no bubbles was detected in the next frame in the circle of radius $30\,\mu$m, we interpreted that the bubble disappeared.
Although few bubbles on the bottom surface were detected in the image analysis, most of the detected bubbles were on the top surface. 

After injection of microbubbles solutions to the capillary, bulk microbubbles approach to the top surface by buoyancy.
Because the glass surface and bubble interface are negatively charged \cite{Uematsu_2020}, the electrostatic repulsion forms a water film at the contact point of the bubble.
When microbubble solutions was produced from deionized water without salts, the film lubricated the bubbles to easily move by a small tilt or flow during the experiments. 
In contrast, using $10\,$mM potassium chloride solutions made the bubbles remain fixed in their position during observation periods.
This can be explained that the addition of salts screens the electrostatic repulsion and makes the film thinner to prevent bubbles from moving. 
As a result, Brownian motion was not observed, and flocculation and coalescence did not occur.
We also assume that the bubbles are not deformed from the spherical shape.
This can be justified that the capillary length of the air/water interface is few millimeters \cite{de_Gennes_2004}, and it is much larger than the microbubble radius in the experiment.  
Taking the spherical shape and thin water film at the contact point of bubbles into account, the contact angle of the bubbles can be considered as 180$^\circ$.
The addition of potassium chloride at $10\,$mM affects the surface tension, Henry constant, and viscosity less than 0.1\%. \cite{Marcus_2015, Battino_1984}
Therefore, we decided to use $10\,$mM potassium chloride solutions in the following experiments.

\section{Results and Discussion}
\subsection{Short summary of the Lifshitz-Slezov-Wagner theory}
Ostwald ripening is a characteristic coarsening process observed in various systems where particles of different sizes are dispersed. 
Because the chemical potentials of molecules in particles depend on their curvatures due to the Gibbs-Thomson effect, the molecules diffuse from small to large particles resulting in the small particles shrinking and the large particles growing. 
The critical radius, which determines the shrinkage or growth, increases with time while the total number of particles decreases. 
This characteristic coarsening was investigated theoretically by Lifshitz and Slezov \cite{Lifshitz1959,Lifshitz1961}, and later by Wagner \cite{Wagner1961}.
In this section, we summarize the LSW theory and then discuss our experimental results.

A multi-particle system is considered where the particle radius is characterized by the radius distribution $F(R,t)$.  
The total number of particles is defined as $N = \int^\infty_0 F(R,t)dR$. 
The continuity equation for $F(R,t)$ is given by
\begin{equation}
	\frac{\partial F}{\partial t} = -\frac{\partial}{\partial R} \left(F\frac{dR}{dt}\right).
\end{equation}
In the LSW theory, the particle radii $R$ are assumed to obey a kinetic equation \cite{Lifshitz1959, Lifshitz1961, Wagner1961} 
expressed by
\begin{equation}
	\frac{dR}{dt} = \frac{K}{{R}^n}\left(\frac{1}{R_\mathrm{c}}-\frac{1}{R}\right),
\label{eq:kinetic}
\end{equation}
where $K$ is the kinetic coefficient, $R_\mathrm{c}$ is the critical radius. 
The index $n$ denotes the types of mass-transfer mechanisms, for example, $n=1$ is volume diffusion, and other $n$ correspond to other mass transfer mechanisms \cite{Alexandrov_2017}
The critical radius $R_\mathrm{c}$ is determined by the conservation of the volume fraction of particles, $\phi$, given by
\begin{equation}
	\phi = \frac{1}{V}\int^\infty_0 \frac{4\pi R^3}{3} F(R,t) dR, 
\label{eq:phi}
\end{equation}
where 
$V$ is the volume of the system. 
For any $n(\ge -1)$, a universal scaled distribution exists \cite{Giron1998,Alexandrov_2017}, and almost any initial distribution of radii approaches it in the limit of $t\to\infty$ \cite{Giron1998,Meerson1999}.
The scaling relations for asymptotic regime are $R_\mathrm{c}\sim t^{1/(n+2)}$ and $N\sim t^{-3/(n+2)}$ where $N$ is the total number of particles.

When the volume diffusion is the mechanism of mass transfer ($n=1$), the critical radius is equal to the mean radius,
\begin{equation}
R_\mathrm{c}=\langle R\rangle = \frac{\int^\infty_0 R F(R,t) dR}{\int^\infty_0 F(R,t) dR},
\end{equation}
and the universal scaled distribution is  
\begin{equation}
P(u) = \frac{ 324 u^2 \mathrm{e}^{-u/(3/2-u)}}{(u+3)^{7/3}(3-2u)^{11/3}}\quad \textrm{for } 0<u<\frac{3}{2},
\label{eq:6}
\end{equation}
otherwise $P(u)=0$ for $u \ge3/2$.
The scaled distribution is defined by $P(u) = \langle R\rangle F(R) / N$ where $u = R/\langle R\rangle$.
When the scaled distribution is described by Eq.~(\ref{eq:6}), the mean radius cubed and mean volume obey the following equations, \cite{Lifshitz1959, Lifshitz1961, Wagner1961, OnukiBook}
\begin{eqnarray}
	\langle R\rangle^3 &=& \frac{4K}{9}t,\label{eq:10}\\
	\langle R^3 \rangle &=& \frac{4A K}{9} t \label{eq:9},
\end{eqnarray} 
where $A = \int^{3/2}_0 u^3 P(u)du=1.13$.
Eq.~(\ref{eq:9}) is usually written as $N = 27 V\phi /16\pi A Kt$, but here we use this description to remove the dependence on $\phi$ \cite{Lifshitz1959, Lifshitz1961}. 

After the seminal works of LSW theory, many experimentalists tested the validity of the theory.   
The experimental scaled radii distribution was generally broader than Eq.~(\ref{eq:6}). 
An explanation was given that Eq.~(\ref{eq:kinetic}) was valid for only infinitesimal particle volume fraction, whereas experiments were usually performed at finite volume fractions. 
To include the effect of finite volume fraction into LSW theory, many theories were suggested \cite{Ardell_1972, Brailsford_1979, Tsumuraya_1983, Marqusee1984, Voorhees1985, Marder1985, Enomoto_1986, Marder1987, Yao_1993, Yao1994, Baldan_2002}. 
These theories reveal that the finite-volume effect broadens the scaled distribution more than the LSW theory and increases the linear coefficients in Eqs.~(\ref{eq:10}) and (\ref{eq:9}).
However, these enhancements of coarsening speed quantitatively differ from each theory 
\cite{Voorhees1985,Baldan_2002} because they were based on different approximations.

\subsection{Time evolution of radius distribution}

\begin{figure}
\includegraphics{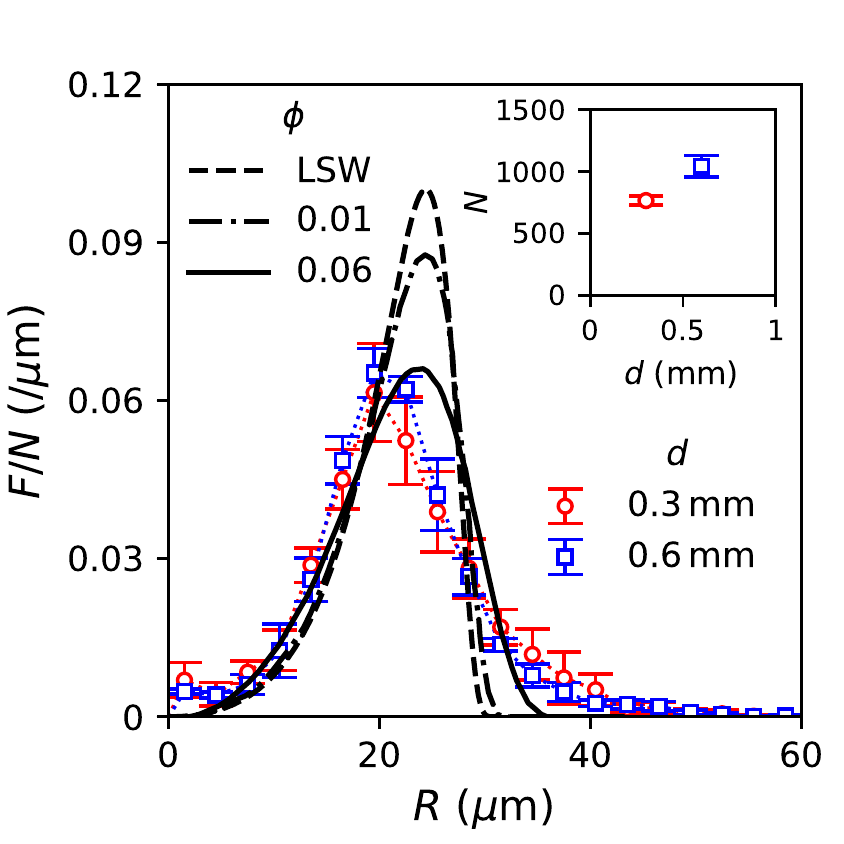}
\caption{
Normalized distributions $F(R,t)/N$ of microbubble radii in the rectangular capillary at $t=1\,$min. 
The discrete histogram $F(R,t)$ is calculated using the bin width $3\,\mu$m.
Average and standard deviation of four independent measurements for each thickness.
The dashed line is the normalized distribution derived by Eq.~(\ref{eq:6}), and 
the dashed-dotted and solid lines are extracted from Ref.~\citenum{Yao_1993} including the effect of finite volume fraction at $\phi = 0.01$ and $0.06$.
$\langle R\rangle =21.5\,\mu$m is plugged in to calculate both theoretical predictions. 
The inset shows the total bubble number in the observed area $S=8.3\,$mm$^2$ with varied thickness. 
}
\label{fig:1}
\end{figure}

In Fig.~\ref{fig:1}, the normalized distributions $F(R,t)/N$ of bubble radii at $t=1\,$min are plotted.
The points are the discrete histograms $F(R,t)$ calculated using bin width of $3\,\mu$m, and $N$ is the total number of bubbles in the observation area. 
The inset shows the total number of observed bubbles, which is not proportional to the capillary thickness $d$ because they are not homogeneously distributed due to buoyancy.
The time $t=1\,$min is the first frame of the observations by microscope after the injection of solution. 
The data points are the average of four independent measurements. 
Fig.~\ref{fig:1} suggests that the normalized distributions at $t=1\,$min are reproducible and independent of the capillary thickness, $d$. 
The mean radius $\langle R \rangle$ is defined by
\begin{equation}
	\langle R \rangle = \frac{1}{N}\sum_{i=1}^{N} R_i,
\end{equation}
where $R_i$ is the radius of the $i$th bubble.
The averages, $\langle R\rangle$, of four independent measurements are $\langle R\rangle=21.5\,\mu$m for $d=0.3\,$mm and $\langle R\rangle= 21.4\,\mu$m for $d=0.6\,$mm, almost identical. 
The dashed line is the universal distribution function for the time $t\to\infty$ derived by LSW theory, Eq.~(\ref{eq:6}).
The experimental radii distribution is broader than the prediction by LSW theory. 
This deviation is not surprising because it has been known that experimental radii distributions of Ostwald ripening are generally broader than that of LSW theory \cite{Voorhees1985, Marder1985,Marder1987,Baldan_2002}.
Many theories considering the effect of finite volume fraction were formulated \cite{Ardell_1972, Brailsford_1979, Tsumuraya_1983, Marqusee1984, Voorhees1985, Marder1985, Enomoto_1986, Marder1987, Yao_1993, Yao1994, Baldan_2002}.
In this work we compare our results to the theory developed by Yao {\it et~al}. \cite{Yao_1993}. 
The dashed-dotted and solid lines are the distribution extracted from the numerical solution in Ref.~\citenum{Yao_1993}. 
These distributions are parameterized by the volume fraction, and $\phi = 0.01$ (dashed-dotted line) and $0.06$ (solid line) \cite{Yao_1993}.
$\phi=0.06$ is the upper limit of the volume fraction applicable to their theory. \cite{Yao_1993} 

\begin{figure}
\includegraphics[width=85mm]{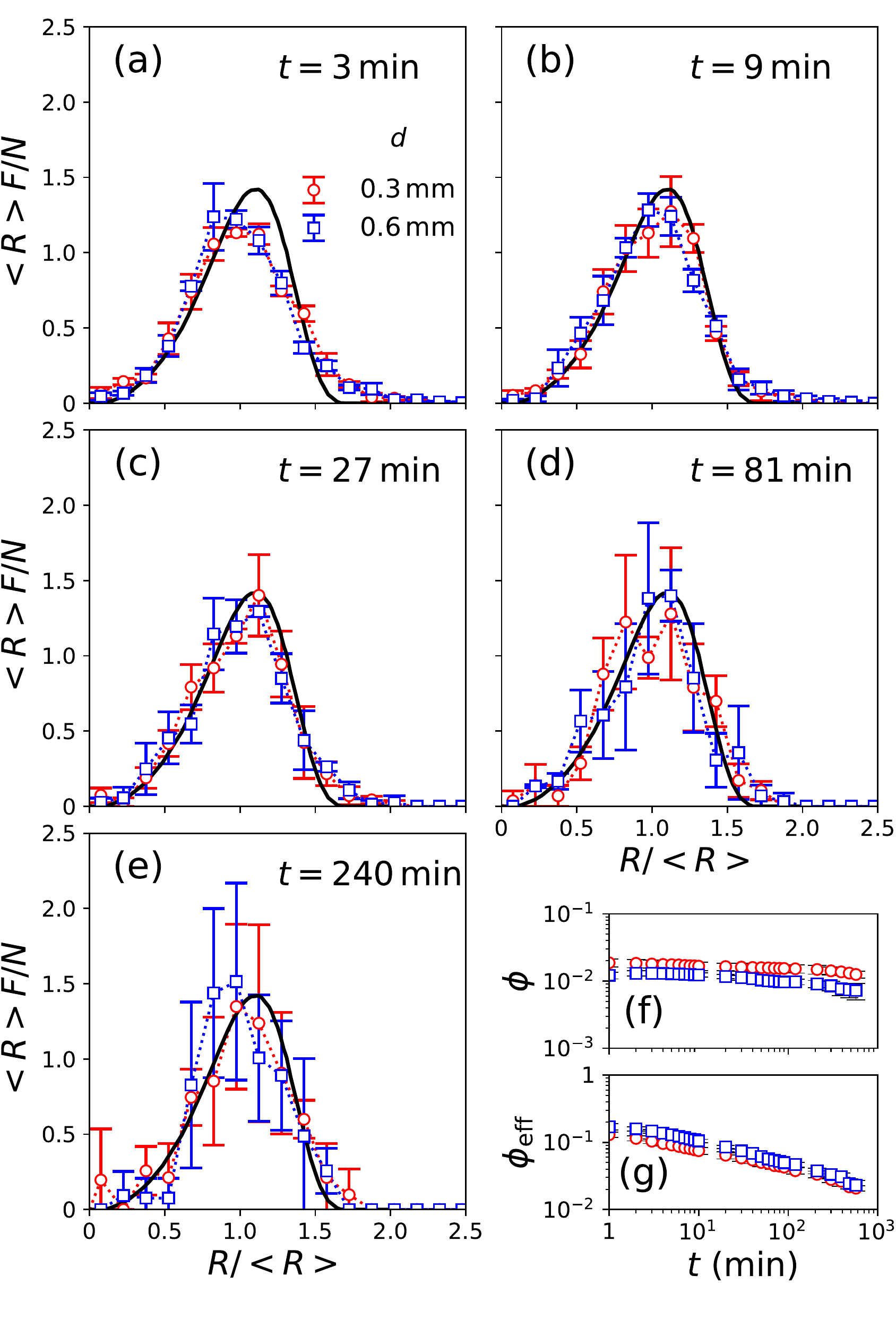}
\caption{
	(a, b, c, d, e) Time evolution of scaled distribution of radii $\langle R\rangle F/N$ is plotted where $0.15$ was used for the bin width. 
	The average and the standard deviation of four independent measurements are plotted for the time series $t=3\,$min, $9\,$min, $27\,$min, $81\,$min, and $240\,$min.
	The solid lines are $P(u)$ derived and extracted from Ref.~\citenum{Yao_1993} for $\phi = 0.06$.
 (f, g) The volume fraction $\phi$ and the effective volume fraction $\phi_\mathrm{eff}$ are plotted as a function of $t$.
}
\label{fig:9}
\end{figure}

The volume fraction of our system is defined by 
\begin{equation}
	\phi = \frac{N}{Sd}\frac{4\pi\langle R^3\rangle}{3},
\end{equation}
where $S$ is the observed area $S=8.3\,$mm$^2$.
It was almost constant with time in agreement with Eq.~(\ref{eq:phi}) and the order of $10^{-2}$ as shown in Fig.~\ref{fig:9}f.
Therefore, if the distribution at $t=1\,$min was already in an asymptotic state, and Yao's theory \cite{Yao_1993} was applicable, the distribution should be close to the dashed-dotted line ($\phi=0.01$) in Fig.~\ref{fig:1}. Instead, however, the observed distribution was close to the solid line ($\phi=0.06$). This quantitative disagreement can be explained by the inhomogeneous position distribution of bubbles caused by buoyancy.
The bubbles were on the glass surfaces and their positions were laterally distributed.
In the middle of the capillary, no bubble exists.
All the theories, including the effect of finite volume fraction, assume that the particles are randomly distributed in a three-dimensional space \cite{Ardell_1972, Brailsford_1979, Tsumuraya_1983, Marqusee1984, Voorhees1985, Marder1985, Enomoto_1986, Marder1987, Yao_1993, Yao1994, Baldan_2002}. 
Therefore, we introduce the effective volume fraction by subtracting the empty region of the capillary, as
\begin{equation}
	\phi_\mathrm{eff} =  \frac{N}{2S\langle R\rangle}\frac{4\pi\langle R^3\rangle}{3}.
\end{equation}
The effective volume fraction $\phi_\mathrm{eff}$ is plotted as a function of time in Fig.~\ref{fig:9}g. As seen, it is no longer constant and the order of $10^{-1}$ in the beginning, is closer to $\phi=0.06$.  
Because the theory does not take the particular lateral distribution of bubbles into account, it is not conclusive. However, the effect of finite volume fraction, in our opinion, explains the broadness of bubble radii distributions in this work.

In Fig.~\ref{fig:9}a, b, c, d, and e, scaled distributions $\langle R\rangle F/N$ at $t=3\,$min, $9\,$min, $27\,$min, $81\,$min, and $240\,$min are plotted together with the distribution considering the finite volume fraction ($\phi=0.06$, solid line) \cite{Yao_1993}.
Figs.~\ref{fig:1} and \ref{fig:9} demonstrated that the scaled distribution was not  significantly different from the first observation at $t=1\,$min.
This suggests that the distribution of radii already achieved a time-independent scaled distribution $1\,$min after the injection.

\subsection{Examination of the radii kinetic equation}

\begin{figure*}
\includegraphics{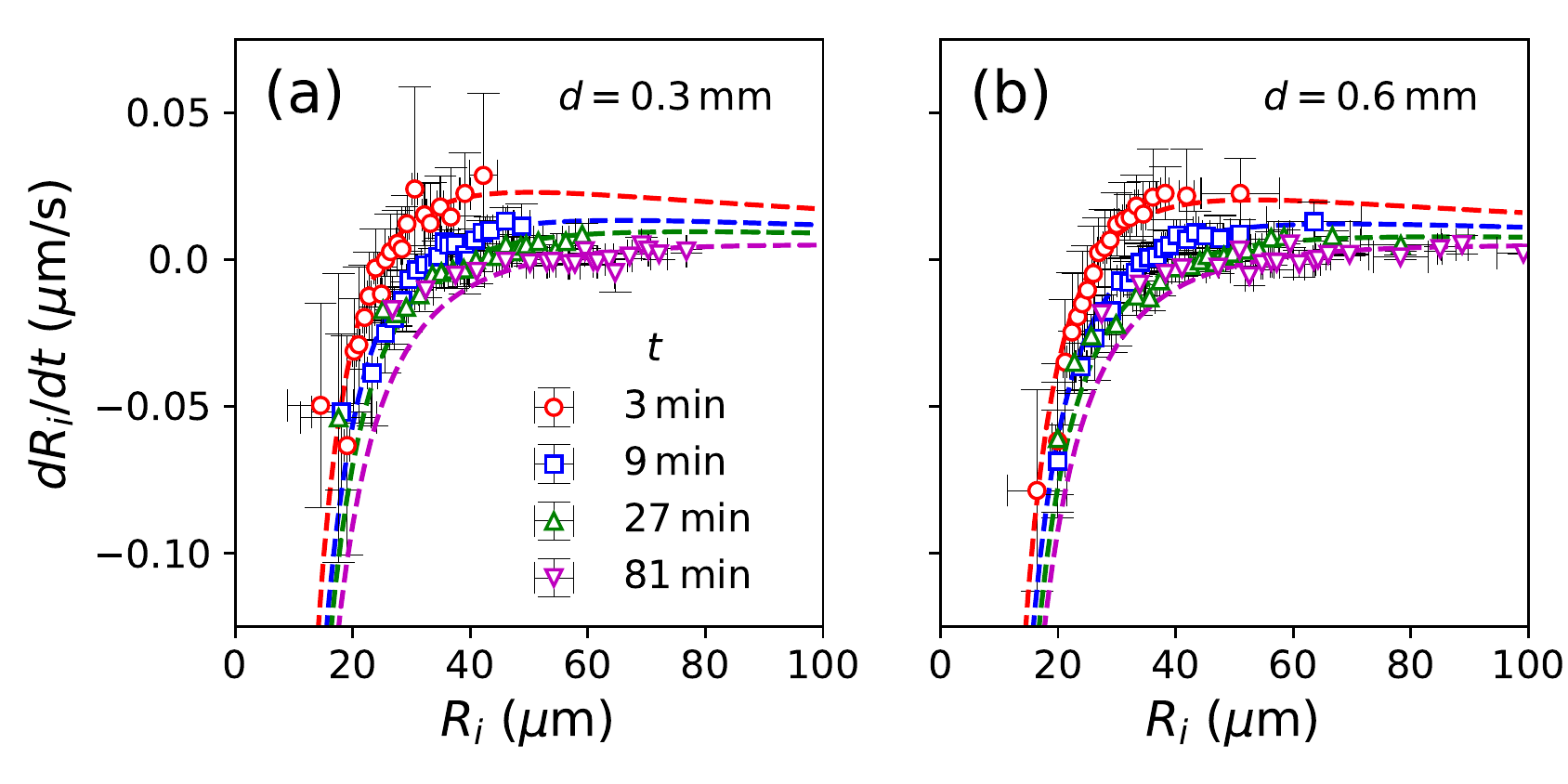}
\caption{
Examination of the kinetic equation of bubble radii.
The left panel (a) denotes the capillary thickness $d=0.3\,$mm whereas the right panel (b) denotes $0.6\,$mm.
$dR_i/dt$ is plotted as a function of $R_i$ at different times: $t=3\,$min (red circle), $9\,$min (blue square), $27\,$min (green up-pointing triangle), and $81\,$min (purple down-pointing triangle). 
Colored dashed lines denote Eq.~(\ref{eq:1}).
The data for $R_i(t+\Delta t)=0$ was omitted. 
All the data points were sorted in $R_i$ and categorized into 20 data groups, and the average and standard deviation were calculated and plotted for each data group.
The data were obtained by a single experimental measurement.
}
\label{fig:3}
\end{figure*}

The late stage of diffusion-limited coarsening dynamics of the supersaturated gas solution is driven by the diffusion of gas molecules between bubbles.
The kinetics of the bubble radii in the mean-field level is given by Eq.~(\ref{eq:kinetic}) with $n=1$ \cite{Slezov2005},
\begin{equation}
\frac{dR_i}{dt} = \frac{K}{R_i}\left(\frac{1}{\langle R\rangle}-\frac{1}{R_i}\right),
\label{eq:1}
\end{equation}
where $K$ is the kinetic coefficient defined by 
\begin{equation}
K = \frac{(2\ln 2)\gamma H D}{p_0}.
\label{eq:K}
\end{equation}
Here, $\gamma$ is the surface tension of the bubble interface, $H$ is the dimensionless Henry constant for air, $D$ is the diffusion constant of air in water, and $p_0=0.10\,$MPa is the atmospheric pressure. 
The derivation of Eq.~(\ref{eq:1}) in the bubble solutions were described in literature \cite{Slezov2005}, assuming that the Laplace pressure of bubbles is much smaller than the atmospheric pressure, $2\gamma/R_i\ll p_0$.
The critical radius $R_\mathrm{c}$ in Eq.~(\ref{eq:1}) is replaced by $\langle R\rangle$ to simplify the comparison with experimental data. 
The factor $\ln 2$ in Eq.~(\ref{eq:K}) becomes 1 when the bubble is isolated in the three-dimensional infinite media \cite{Slezov2005}.
We assumed that the contact angle of bubbles on the glass surface is $180^\circ$ and bubbles are not deformed. 
In that case,  the factor, $\ln 2$, can be obtained by the method of image charges in electrostatics \cite{Liebermann1957,Kentish_2006}.
Table~\ref{tab:param} lists the literature values of $\gamma$, $H$, and $D$.

\begin{table}
\caption{Surface tension of the air/water interface, and Henry and diffusion constant of air used in this paper.}
\label{tab:param}
\vspace{2mm}
\begin{tabular}{cccc}
\hline\hline
Name & Quantity & Value & Ref.\\\hline
Surface tension & $\gamma$ & $72\,$mN/m & \cite{CRC2016}\\ 
	Henry constant &$H$ & $1.9\times 10^{-2}$ & \cite{CRC2016,Sander2015}\footnote{Calculated by $H=1353.96(0.2H^{xp}_\mathrm{O_2}+0.8H^{xp}_\mathrm{N_2})$ where $H^{xp}_\mathrm{O_2} = 2.293\times 10^{-5}\,$/atm and $H^{xp}_\mathrm{N_2}=1.183\times 10^{-5}\,$/atm are the Henry constants of oxygen and nitrogen gas at $298.15\,$K defined by the molar fraction divided by the pressure \cite{CRC2016}, and $1353.96\,$atm is the conversion constant for non-dimensional Henry constants \cite{Sander2015}.}\\ 
Diffusion constant & $D$ & $2.1\times 10^{-9}\,$m$^2$/s & \cite{CRC2016}\footnote{Calculated by $D=0.2D_\mathrm{O_2}+0.8D_\mathrm{N_2}$ where $D_\mathrm{O_2} = 2.42\times 10^{-9}\,$m$^2$/s and $D_\mathrm{N_2}=2.0\times 10^{-9}\,$m$^2$/s are the diffusion constants of oxygen and nitrogen gas in water at $298.15\,$K \cite{CRC2016}.}\\
\hline\hline
\end{tabular}
\vspace{2mm}
\end{table}

First, we examined the kinetic equation of the diffusion-controlled Ostwald ripening, Eq.~(\ref{eq:1}).
In Fig.~\ref{fig:3}, experimentally obtained  $dR_i/dt$ were plotted as a function of $R_i$ at different times $t=3\,$min (red circle), $9\,$min (blue square), and $27\,$min (green up-pointing triangle), and $81\,$min (purple down-pointing triangle). 
For the discretization,
\begin{equation}
\frac{dR_i}{dt} = \frac{R_i(t+\Delta t) - R_i(t-\Delta t)}{2\Delta t},
	\label{eq:3}
\end{equation}
was used where $\Delta t = 1\,$min.
Because the number of data points is as large as $10^2$ to $10^3$, all the data points were sorted in $R_i$ and categorized into 20 data groups, and the average and standard deviation were calculated and plotted for each data group.
The colored dashed lines denotes Eq.~(\ref{eq:1}), which agree well with the experimental data.
Surprisingly, this comparison needs no fitting parameters, demonstrating that Eq.~(\ref{eq:1}) well works as the mean field description of the bubble radius kinetics.
Furthermore, this quantitative agreement was not obtained without the factor $\ln 2$ in Eq.~(\ref{eq:K}).
This result implies that the inhomogeneous distribution of the bubble positions is not relevant for the mean field description because Eq.~(\ref{eq:1}) assumes infinite separation between any bubble pairs. 

For small bubbles about $R<10\,\mu$m, Fig.~\ref{fig:3} shows no data.
The reason is that for small bubbles, the shrinkage speed is fast, and they often disappear during the period $2\Delta t$. 
In the calculation of shrinkage speed in Fig.~\ref{fig:3}, we rejected the data corresponding to bubbles disappearing at $t+\Delta t$.

\begin{figure}[t]
	\includegraphics[width=85mm]{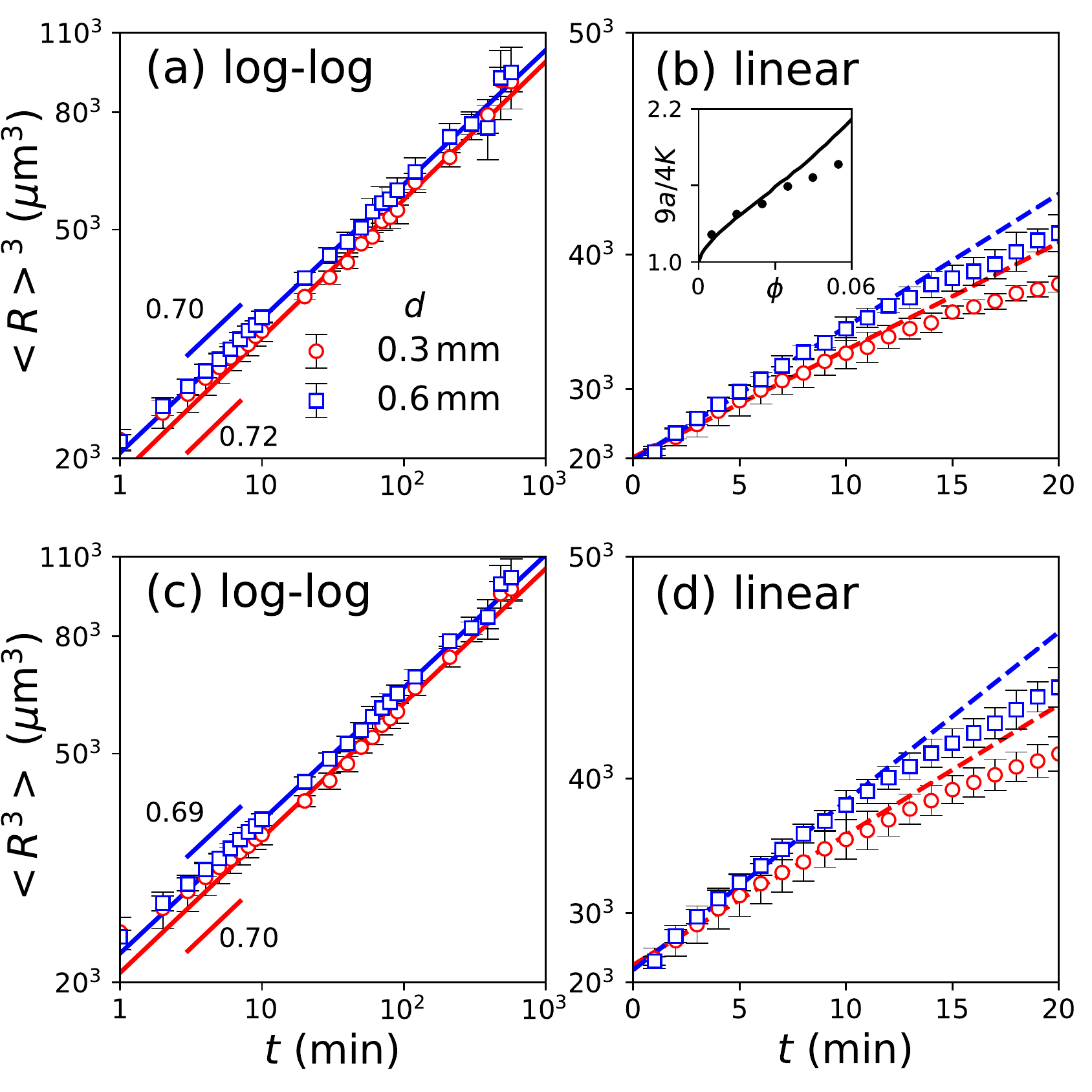}
\caption{
	$\langle R\rangle^3$ and $\langle R^3\rangle$ as a function of $t$ in double logarithmic and linear plots.
(a) $\langle R\rangle^3$ in double logarithmic scale.
(b) $\langle R\rangle^3$ in linear scale.
(c) $\langle R^3 \rangle$ in double logarithmic scale.
(d) $\langle R^3 \rangle$ in linear scale.
The points and error bars are the averages and standard deviations of four independent measurements.
In (a) and (c), the entire range of data was used for the fit (colored solid lines), whereas in (b) and (d), the data of $t\le10\,$min was used for the fit (colored dashed lines).
The numbers in (a) and (c) denote the exponents of fitted power law functions.
	The inset in (b) shows the theoretical and simulated normalized coarsening speed $9a/4K$ as a function of the volume fraction $\phi$ extracted digitally from Ref.~\citenum{Yao_1993}.
	Line represent theoretical calculation, while the points are their simulation results \cite{Yao_1993}. 
	}
\label{fig:2}
\end{figure}

\subsection{Examination of coarsening speed}

Although our scaled distributions shown in Figs.~\ref{fig:1} and \ref{fig:9} have a different shape compared to that predicted by the LSW theory, it is worth discussing the experimental data using Eqs.~(\ref{eq:10}) and (\ref{eq:9}). 
In Fig.~\ref{fig:2}a and c, the mean radius cubed, $\langle R\rangle^3$, and the mean volume $\langle R^3\rangle$ are plotted as a function of $t$ in a double logarithmic scale.  
Both $\langle R\rangle^3$ and $\langle R^3\rangle$ were monotonically increasing and well fitted to the power law $t^\alpha$ (colored solid line), and the exponents of $\langle R\rangle^3$ and $\langle R^3\rangle$ are similar and almost independent of $d$. 
When the general version of kinetic equation, Eq.~(\ref{eq:kinetic}), is taken into account, these exponents about $\alpha=0.7$ correspond to the intermediate index between $n=3$ and $4$. 
However, these higher-order kinetic equations are derived for other mass transfer mechanisms in supersaturated solid solutions \cite{Alexandrov_2017}. Thus, it is unsuitable for bubble/water systems where volume diffusion is the primary mechanism for mass transfer.
In our opinion, the apparent power-law behavior and dependence of power-law amplitude on $d$ are caused by the effect of finite volume fraction.

\begin{table}
	\caption{Fitted and theoretical values for Eqs.~(\ref{eq:10}) and (\ref{eq:9}).} 
\label{tab:2}
\vspace{2mm}
\begin{tabular}{cccc}
\hline\hline
				$d$ & $0.3\,$mm	& $0.6\,$mm & LSW \\\hline
	$a$ ($10^3\,\mu\mathrm{m}^3/$min) 	&3.0		&3.7		&1.1 \\
	$c_1$ ($\mu$m) 			&20 		&20 		&--\\
	$b$ ($10^3\,\mu\mathrm{m}^3/$min)		&3.6	&4.6	&1.2 \\
	$c_2$ ($\mu$m)  &23& 23 &-- \\
	\hline\hline
\end{tabular}

\vspace{2mm}
\end{table}

In Figs.~\ref{fig:2}b and d, $\langle R\rangle^3$ and $\langle R^3\rangle$ were plotted using a linear scale up to $t=20\,$min. 
These plots revealed that both  $\langle R\rangle^3$ and $\langle R^3\rangle$  grow linearly in the first ten minutes, and their slopes  decrease with time. 
The colored dashed lines are the linear fits, $\langle R\rangle^3=at+{c_1}^3$ and $\langle R^3\rangle=bt+{c_2}^3$, using the data of $t \le  10\,$min. 
Because Eqs.~(\ref{eq:10}) and (\ref{eq:9}) are asymptotic equations for $t\to\infty$, addition of an arbitrary constant does not contradict them.
	Table~\ref{tab:2} summarizes the fitted values of $a$, $b$, $c_1$, and $c_2$ and predictions by the LSW theory.
	The fitted $a$ and $b$ are a few times larger than the LSW predictions.
	This is possibly because the effect of finite volume fraction increases the coarsening speed, $a$ and $b$, from the values of the LSW theory.
	When $\phi = 4\pi N\langle R^3\rangle /3Sd$ shown in Fig.~\ref{fig:9}f is used as the volume fraction, 
	the volume fraction of the capillary $d=0.3\,$mm becomes larger than that of $d=0.6\,$mm because the empty region in the middle is included in the calculation.  
	As shown in the inset of Fig.~\ref{fig:2}b, the normalized coarsening speed $9a/4K$ is an increasing function of the volume fraction \cite{Yao_1993}.
	Therefore, $\phi$ is not a good variable to characterize the effect of finite volume fraction in our system, and we use the effective volume fraction $\phi_\mathrm{eff}$ shown in Fig.~\ref{fig:9}g. 
	In the beginning, $\phi_\mathrm{eff}$ is the order of $10^{-1}$, whereas it decreases later to the order of $10^{-2}$. 
	The normalized coarsening speed $9a/4K$ in Fig.~\ref{fig:2}b is $2.8$ for $d=0.3\,$mm and $3.4$ for $0.6\,$mm, and then these  decrease with time.
	These behaviors can be semi-quantitatively explained by the time dependence of $\phi_\mathrm{eff}$ and the $\phi$ dependence of the coarsening speed as follows. 
	In the beginning, the relatively high volume fraction, $\phi_\mathrm{eff}\sim 10^{-1}$, enhanced the speed of coarsening to about three times of the LSW prediction, and later the relatively low volume fraction $\phi_\mathrm{eff}\sim 10^{-2}$ slowed down the coarsening speed.
	As a result, the range of linear growth was limited for the first ten minutes. 
	Assuming that the coarsening speed is affected by the effective volume fraction as $a\sim\phi_\mathrm{eff}$, equating $a\sim \phi_\mathrm{eff}\sim \langle R\rangle^{-1}\sim t^{-\alpha/3}$ and $a\sim \langle R\rangle^3/t\sim t^{\alpha-1}$ yields $\alpha =0.75$, which is very close to our fitted exponents of about $\alpha=0.7$. 
	Because a theory for randomly distributed coarsening particles in two-dimensional space is lacking, we cannot compare the coarsening speed with a more suitable theory.
	Ostwald ripening of organic liquid droplets in water exhibited the linear evolution of $\langle R \rangle^3$ for the full range of the observation period of about $100\,$min \cite{Kabalnov1987}.
	This can be explained by the small coarsening speed of that system of about $10^{-1}\,\mu$m$^3$/min.

\subsection{Bubble shrinkage dynamics}

\begin{figure}
	\includegraphics[width=85mm]{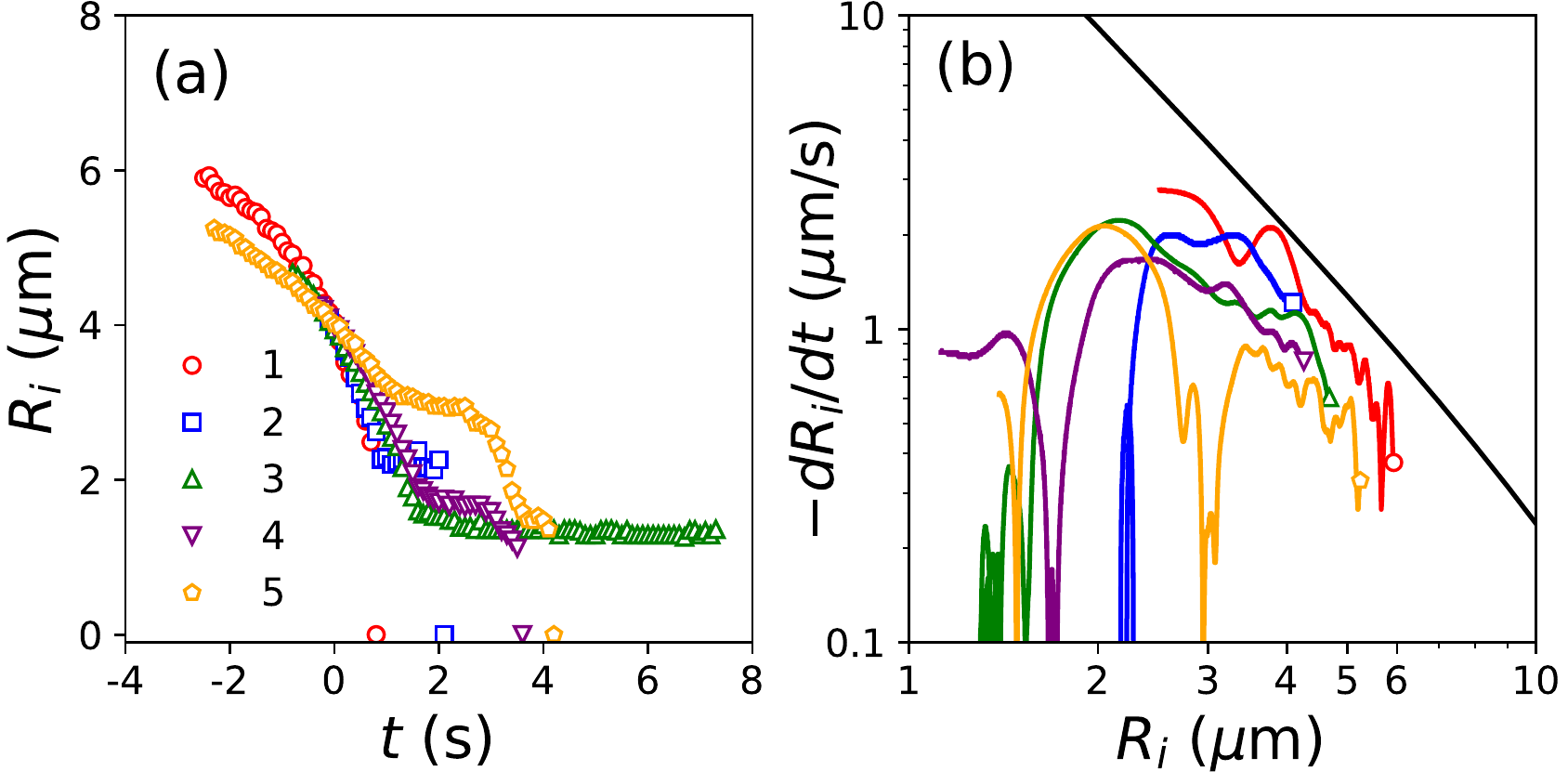}
\caption{
	(a) Radii of small shrinking bubbles as a function of time. 
	The reference time $t=0$ is set at radius of each bubble by $4\,\mu$m.
	(b) Shrinkage speed $-dR_i/dt$ of individual bubbles as a function of $R_i$ plotted by the colored solid lines. 
	Raw data plotted in (a) were interpolated by spline curves, and $-dR_i/dt$ was obtained by differentiation of the spline curves.
	The black solid line is Eq.~(\ref{eq:1}) where $\langle R \rangle =25\,\mu$m was plugged in. 
}
\label{fig:5}
\end{figure}

In this section, shrinkage dynamics of small bubbles in a bubble solution is investigated.
The shrinkage dynamics of a single bubble in water has been studied using the Epstein-Plesset equation so far \cite{Epstein1950,Liebermann1957,Duncan_2004,Kentish_2006}, but not yet in a bubble solution to the best of our knowledge.
Fig.~\ref{fig:5}a shows the radius of shrinking bubbles as a function of time. 
The images were taken by an objective lens $\times 20$ at $10\,$fps, and the capillary thickness was $d=0.3\,$mm.
The approximate time after the injection is $2\,$min, and the mean radius is about $25\,\mu$m, which is much larger than the radii of shrinking bubbles. 
The reference time $t=0$ here is set at the time of $R_i=4\,\mu$m. 
These experimental data exhibit anomalous shrinkage, which cannot be   described by Eq.~(\ref{eq:1}), for example, the slowdown of the shrinkage speed and pinning of the radius.
A bubble, number 3 (green up-pointing triangle), was stable during the observation period at the radius $R_i=1.3\,\mu$m.
A bubble, number 1 (red circle) shrank and disappeared as expected of Eq.~(\ref{eq:1}), whereas other bubbles, number 2, 4, and 5 (blue square, purple down-pointing triangle, and orange pentagon), exhibited the slowdown of shrinkage and pinning of the radius for a few seconds.  
Although most of bubbles disappeared ($R=0$) in the last frame in Fig.~\ref{fig:5}a, it was not clear whether they actually disappeared or just escaped detection by the microscope.  

Fig.~\ref{fig:5}b shows the speed of shrinkage $-dR_i/dt$ of individual bubbles as a function of $R_i$.
The colored lines in Fig.~\ref{fig:5}b are calculated by the numerical differentiation of the spline interpolation of the discrete data in (a).
Therefore, small windings of the lines in (b) are artifacts.
The black solid line represents Eq.~(\ref{eq:1}) with $\langle R\rangle=25\,\mu$m, which shows somewhat  larger values than the experimental observations for $R>4\,\mu$m.
However, when the radius was smaller than $4\,\mu$m, the shrinkage speed of some bubbles (number 2, 3, 4 and 5) was significantly slowed down, and then, the radius was pinned for a few seconds.
The reason for these anomalous shrinkages is not clear. 
Similar slowdown was reported in a shrinking bubble in partially degassed water \cite{Liebermann1957}. 
Furthermore, this anomalous shrinkage is not deterministic because we could not find any principle explaining it, implying that this is controlled by uncontrolled factors such as the cleanness of the glass surface and impurity at the bubble interfaces. 

We speculate two scenarios for these anomalous shrinkage.
The first scenario is that the thin water film at the contact point between the bubble and glass surface was ruptured by hydrophobic impurities on the glass surface. 
The contact line dynamics is generally slower than the diffusion of gas molecules.
Therefore, when the size of dry contact area is few micrometers, the shrinkage dynamics will be slow and possibly pinned at this length scale.

The second scenario is based on a mechanism of stabilizing nanobubbles proposed recently \cite{Satpute_2021}. 
During the shrinkage of a microbubble with few tens micrometer radius, the absorbed ions at the interface are concentrated, and the resultant electrostatic  repulsion will reduce the shrinkage speed than the prediction of Eq.~(\ref{eq:1}) \cite{Satpute_2021}.
Restarting the shrinkage would be caused by desorption of the adsorbed ions. 
We do not think that the adsorbed ions are hydroxide ions but charged impurities because the adsorption of hydroxide ion contradicts the surface tension measurement of basic electrolyte solutions \cite{Uematsu_2017,Uematsu_2020}.

Although these anomalous shrinkage behaviors must be important for the stability of long-lived nanobubbles, further investigation regarding this is beyond the scope of this paper because the paper clarifies that microbubble solutions averagely undergo diffusion-limited Ostwald ripening.

\section{conclusion}

We performed experiments on the kinetics of radii in aqueous microbubble solutions confined in a rectangular glass capillary and examined the theories of Ostwald ripening using experimental data.   
The images of the bubbles were analyzed to obtain the radii and total number of bubbles, and the distribution function of bubble radii was calculated. 
The calculated radii distribution is closer to the universal distribution with the effect of finite volume fraction included, rather than that derived from the LSW theory, and the scaled distribution does not change. 
We found that the growth and shrinkage speed of each bubble are averagely governed by diffusion-limited Ostwald ripening.
The kinetic coefficient in the kinetic equation is calculated by the surface tension, Henry constant, diffusion coefficient of gas in literature, and the numerical factor representing the binding of bubbles on the glass surface. We find a quantitative agreement of our experimental data without any fitting parameters, suggesting that the inhomogeneous distribution of the bubble positions is not relevant for the mean-field description. This demonstrates that the coarsening of bubbles is driven by diffusion-limited Ostwald ripening.

Furthermore, we revealed that the mean radius cubed and mean volume exhibit a linear time evolution. The coefficients are slightly larger but in the same order of magnitude as that predicted by LSW theory. 
The coarsening speed is enhanced due to the effect of finite volume fraction. 
Because the coarsening speed is theoretically calculated to be an increasing function of the volume fraction \cite{Ardell_1972, Brailsford_1979, Tsumuraya_1983, Marqusee1984, Voorhees1985, Marder1985, Enomoto_1986, Marder1987, Yao_1993, Yao1994, Baldan_2002},
the effective volume fraction where $d$ is replaced by $2\langle R\rangle$ is more reasonable to characterize the finite volume effect in the two-dimensional distributed bubble system. 
However, a theory of Ostwald ripening for such a special system with a lateral particle position distribution has not been constructed, limiting a quantitative comparison between theory and experiments.

Lastly, the decrease in the shrinkage speed and pinning of the radii of a small microbubble in a microbubble solution are discussed in detail.
These kinetics cannot be described by the mean-field kinetic equation of Ostwald ripening. Instead, we consider that they are influenced by uncontrolled factors such as the cleanness of the glass surface and impurity at the bubble interfaces.
However, on average, these anomalous behaviors have minor effects on the overall coarsening.

Bubbles/water system is unique to study the Oswald ripening because all the relevant physicochemical properties such as the surface tension, Henry constant, and diffusion constant are quantitatively known in literature. 
It serves as a model experimental system for studying Ostwald ripening when an improved theory includes the lateral distribution of their positions.
In addition, this study is the first step to understanding the stability of nano- and microbubble solutions from a many-body effect perspective. 
Application of this picture to bulk nanobubble solutions will be the focus of our future work. 
We believe this paper will elucidate fundamental properties of bubble solutions and stimulate further technological applications.

\section*{Acknowledgements}

This work was supported by JST, PRESTO Grant number JPMJPR21O2, JSPS KAKENHI Grants 18KK0151 and 20K14430, a grant from the Kurita Water and Environment Foundation (21E006), and support from the Qdai-jump Research Program from Kyushu University (R3-01302).
Y.U. thanks Kenichiro Koga, Toshihiko Eguchi, and Akimi Serizawa for valuable discussions. 

\section*{Data Availability Statements}
The data that support the findings of this study are available from the corresponding author upon reasonable request.

\bibliography{droplet}
\end{document}